\documentclass[aip,reprint,apl]{revtex4-1}
\usepackage{graphicx}

\begin{document}

\title{Synthesis of photochromic oxygen-containing yttrium hydride}

\author{J. Montero}
\thanks{jose.montero.amenedo@ife.no}
\affiliation{Institute for Energy Technology, P.O. Box 40, NO-2027 Kjeller (Norway)}
\author{F. A. Martinsen}
\affiliation{Institute for Energy Technology, P.O. Box 40, NO-2027 Kjeller (Norway)}
\author{M. Lelis}
\affiliation{Centre for Hydrogen Energy Technologies, Lithuanian Energy Institute, 3 Breslaujos st., Kaunas (Lithuania)}
\author{S. Zh. Karazhanov}
\affiliation{Institute for Energy Technology, P.O. Box 40, NO-2027 Kjeller (Norway)}
\author{B. C.  Hauback}
\affiliation{Institute for Energy Technology, P.O. Box 40, NO-2027 Kjeller (Norway)}
\author{E. S. Marstein}
\affiliation{Institute for Energy Technology, P.O. Box 40, NO-2027 Kjeller (Norway)}

 \date{\today}

\begin{abstract}
Photochromic oxygen-containing yttrium hydride has been synthesized using a two step process. The process consists of an initial sputter deposition of  oxygen-free yttrium hydride YH$_\mathrm{x}$, followed by a controlled reaction with air that causes incorporation of oxygen into the material. An \emph{in-situ} study of the YH$_\mathrm{x}$ as it reacted with oxygen was made possible by applying an aluminium capping layer with a low but non-zero oxygen permeability prior to air exposure. During the reaction, the visual appearance of the YH$_\mathrm{x}$ went from  dark and opaque to transparent and yellowish. This transition was accredited incorporation of oxygen into the material, as shown through analysis with energy dispersive x-ray spectroscopy (EDS) and x-ray photoemission spectroscopy (XPS). Grazing incidence x-ray diffraction (GIXRD) studies revealed an \emph{fcc} structure both before and after oxygen exposure, with a lattice parameter that increased from 5.2 \AA \  to 5.4 \AA \ during the reaction. With the lattice parameter in the oxygen free YH$_\mathrm{x}$ being equal to that reported earlier for YH$_\mathrm{2}$, our findings suggest that the non-reacted YH$_\mathrm{x}$ in this study is in-fact YH$_\mathrm{2}$.
\end{abstract}


\maketitle 

Smart windows based on chromogenic materials have in recent years emerged as promising candidates for the next generation of energy-efficient windows.\cite{YeMengLongEtAl2013}  One group of chromogenic materials is comprised of photochromic compounds, which are characterized by the ability to change their optical properties reversibly when exposed to light, and thus can be used for the fabrication of windows with dynamic and situation-determined solar control properties.\cite{smith2013green} Photochromic materials exist both in organic and inorganic varieties,\cite{Towns2016} where some inorganic examples include titanium oxide, tungsten oxide, molybdenum oxide and yttrium hydride.\citep{Mongstad1}

The discovery of photochromic properties in yttrium hydride came as a result of the increased scientific activity\citep{DerMolen1,Dam1,Gogh1,Lokhorst1,Kerssemakers1,Richardson1} that followed  the discovery  of the switchable optical properties in yttrium and lanthanum-based hydrides.\citep{Huiberts1} First, photochromism in yttrium hydride was observed only at high pressures,\citep{Ohmura1} but by incorporating oxygen into the material,  Mongstad $\textit{et al.}$ \citep{Mongstad1} later showed that photochromic properties also could be obtained at ambient conditions. 

The material in the latter study was named \it oxygen-containing yttrium hydride \rm (YH$\mathrm{_x}$:O), as it was experimentally shown to contain up to $\sim$ 30 \% at. oxygen.\citep{Mongstad2} The oxygen in YH$\mathrm{_x}$:O has earlier been reported to unintentionally get incorporated during the reactive sputter deposition process used for fabrication, where the low electro-negativity of yttrium was believed to cause a reaction with oxygen-containing impurities in the deposition chamber.\citep{Mongstad1,Mongstad2,You1} YH$\mathrm{_x}$:O have been found to exhibit an \emph{fcc} cubic structure similar to the one of YH$_2$ but with larger lattice parameter ($\sim 5.4$ \AA).\cite{Maehlen1} Its band gap has been measured as $\sim 3.0$ eV, \cite{You1} making YH$\mathrm{_x}$:O thin films very transparent in their clear state. 

A recursive problem with the aforementioned reports on YH$\mathrm{_x}$:O is that they fail to supply a definite explanation for how oxygen gets introduced into the material. It has been proposed that oxygen originate from impurities within the sputter deposition chamber used for fabrication, \citep{Mongstad1,Mongstad2,You1} but experimental observations have this far been found to contradict this claim. Films with various electronic properties, oxygen contents and optical appearances have for example been fabricated within the same deposition chamber with a constant impurity content, \citep{Mongstad1,Mongstad2,Mongstad3,You1,You2,Maehlen1}  hence suggesting that the source of oxygen might be a different one. 

In this work, we report on a method for how to synthesise photochromic YH$\mathrm{_x}$:O,  while presenting a definite proof of the source of oxygen. The process consists of an initial reactive magnetron sputter deposition of oxygen free yttrium hydride films, followed by an exposure to air that ensured incorporation of oxygen into the material. All films were inspected optically and analysed using x-ray diffraction (XRD), energy dispersive x-ray spectroscopy (EDS) and x-ray photo-electron spectroscopy (XPS).

Oxygen free yttrium hydride thin films (YH$_{\mathrm{x}}$) of $\sim$ 500 nm thickness were initially deposited on soda lime glass through sputtering of metallic yttrium in a hydrogen-containing atmosphere. A 99.99\% pure metallic yttrium target was used and the sputter process was performed in a Leybold Optics sputter deposition unit operated with a power density of 1.33 W/cm$^2$. The base pressure in the chamber was $\sim$ 10$^{-6}$ mbar and the deposition was performed at $10^{-2}$ mbar in an approximate 1:10 flux ratio of hydrogen (5N) and argon (5N). No oxygen was intentionally introduced in the deposition chamber. Prior to removal from the vacuum chamber, some YH$_{\mathrm{x}}$ films were capped with $\sim$ 200 nm of aluminium. The purpose of this capping was to protect the yttrium hydride films from oxidation upon subsequent exposure to air. In order to minimize the possibility of any reaction with oxygen in the period between fabrication and characterization, these films were stored in sealed containers purged with nitrogen.

The aluminium capping was fabricated with a low but non-zero oxygen permeability. Thus upon exposing the combined film to air, the aluminium ensured a limited flow of oxygen to the YH$_{\mathrm{x}}$ that allowed for an \emph{in-situ} study of the reaction that followed. The visual development of the films during the reaction was studied optically from the substrate side of the sample, while the crystal structure was monitored through periodic grazing incidence X-ray diffraction (GIXRD) measurements. In the latter, a measurement was performed every $\sim$ 30 minutes employing a Bruker Siemens D5000 diffractometer, operated with a 2$^{\circ}$ grazing incidence angle of Cu-K$_{\alpha}$ radiation. The optical reflectance of the capped films was measured before and after the complete reaction and compared with that of a non-capped, reacted film. These measurements were conducted from the substrate side of the films and performed over the 300-1700 nm wavelength range, employing two Ocean Optics spectrophotometers (QE65000 and NIRQUEST512) equipped with an integrating sphere.

The aluminium capped YH$_{\mathrm{x}}$ thin films were characterized using a Hitachi S-4800 scanning electron microscope (SEM) mounted with a Thermo Noran energy dispersive x-ray spectroscope (EDS). Secondary electron imaging (SEI) was mainly used for investigation of the film integrity, while EDS was employed in order to obtain an estimate of the oxygen concentration in the yttrium hydride. In order to get a sufficient signal from the capped YH$_{\mathrm{x}}$ films, the acceleration voltage was kept at 15 keV during EDS-measurements. This enabled the beam to penetrate through the capping layer and generate a signal from the YH$_{\mathrm{x}}$. Since hydrogen is not detectable with EDS, all oxygen concentrations (C$_{\mathrm{O}}$[\%]) are given relative to the concentration of yttrium only: C$_{\mathrm{O}}$[\%]=C$_{\mathrm{O}}$/(C$_{\mathrm{O}}$+C$_{\mathrm{Y}}$). 

X-ray photo-electron (XPS) measurements were performed using a PHI 5000 Versaprobe XPS unit in order to verify the oxygen concentrations obtained using EDS. In the case of the capped films, an analysis was performed both on as-deposited and on reacted films. This was done by sputtering with a 3 keV argon gun through the aluminium capping before a scan was executed. The uncapped samples were analysed after pre-sputtering of the surface.

\begin{figure}[hptb]
\centering
\includegraphics[width=0.95\linewidth]{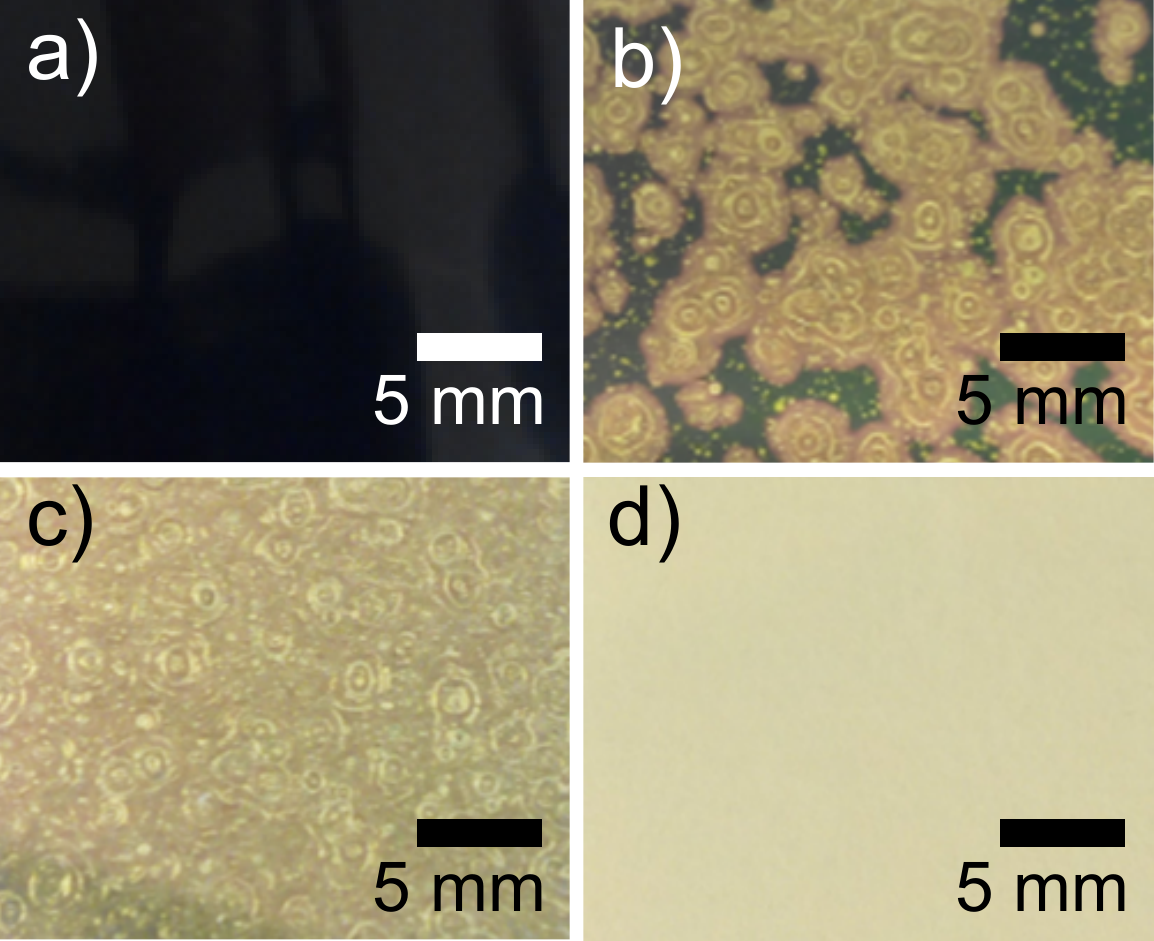}
\caption{Photographs showing samples of aluminium capped (a-c) and uncapped (d) YH$_{\mathrm{x}}$ thin films depicted from the substrate side of the sample. Immediately after removal from the vacuum chamber the aluminium capped samples had a dark opaque appearance (a), which upon exposure to air gradually turned light yellow and transparent (b).  Eventually, the reaction reached an equilibrium (c), resulting in a film not unlike those obtained through fabrication without a capping (d).}
\label{fig:Oxidation2}
\end{figure}

Upon immediate removal from the vacuum chamber, the aluminium capped YH$_{\mathrm{x}}$  films were observed to have a dark, opaque appearance (Fig. \ref{fig:Oxidation2}a). This differed from non-capped samples fabricated under the same conditions, which likely due to an instantaneous reaction with air, were observed to have a transparent, light yellow appearance (Fig. \ref{fig:Oxidation2}d). 
After a few minutes of air exposure, however, the capped samples were observed to have begun undergoing a slight reaction.  Small parts of the films were observed to increasingly transform into a transparent form (Fig. \ref{fig:Oxidation2}b), not unlike the non-capped films. Upon prolonged storage in air ($\sim$ hours), this reaction spread across the respective films, eventually causing the entire film to transform (Fig. \ref{fig:Oxidation2}c). 

\begin{figure}
\centering
\includegraphics[width=0.99\linewidth]{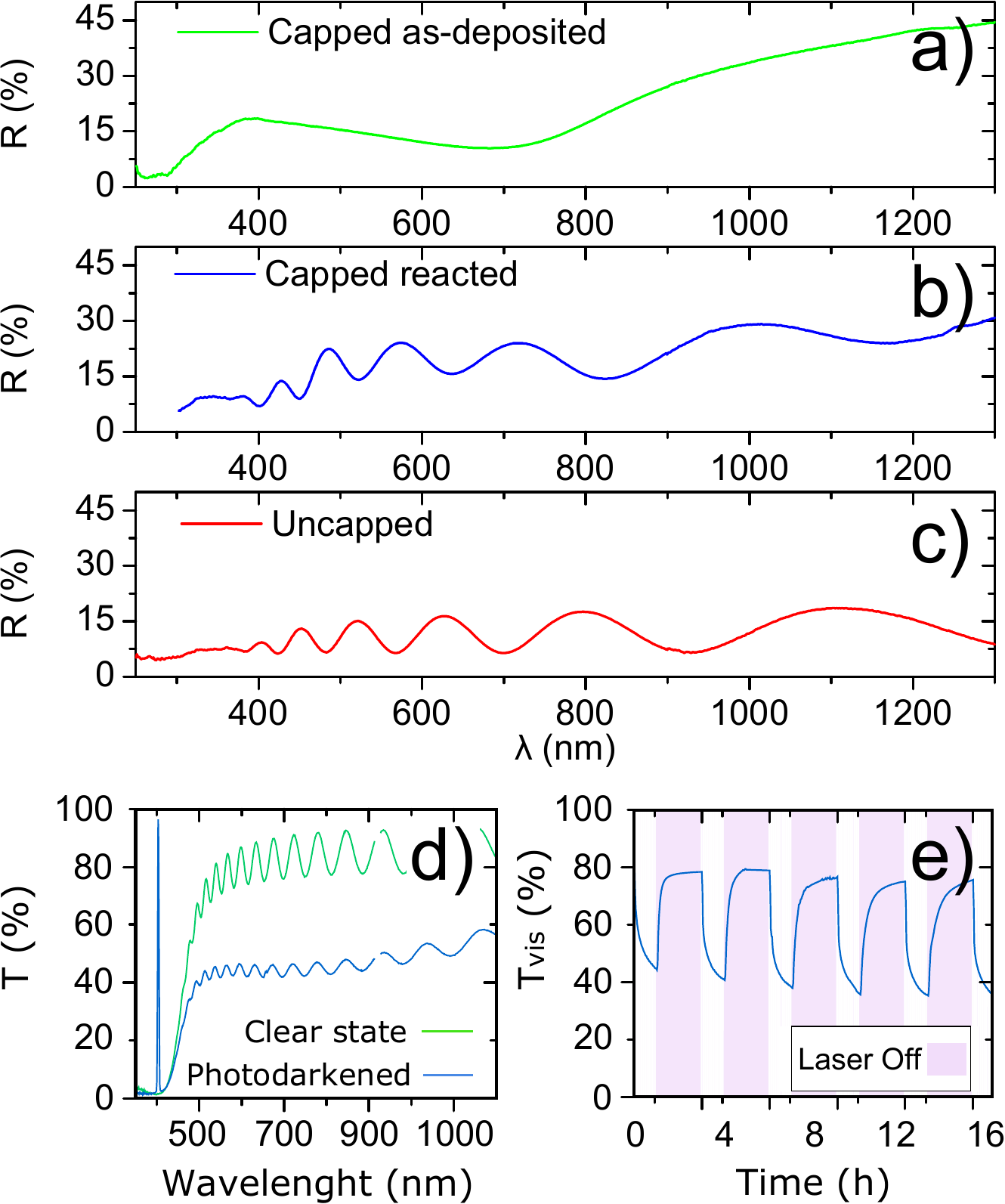}
\caption{The reflectance from the substrate side for an aluminium capped yttrium hydride thin film before (a) and after (b) its reaction with air, as well as for an uncapped, reacted film (c). The spectral photochromic response of an uncapped, reacted film is visualized in (d) together with its repeated cycleability averaged over the 500-800 nm range (e).}
\label{fig:Fig_Optics}
\end{figure}

The reflectance of the films presented in Figure \ref{fig:Oxidation2}a, c and d, is shown in Figure \ref{fig:Fig_Optics}. The capped unreacted YH$_\mathrm{x}$ film (Fig. \ref{fig:Fig_Optics} a) shows a high reflectance for long wavelengths, a property that indicates the presence of free charge carriers.\citep{ginley2010handbook} The lack of optical interferences in the reflectance implies that the unreacted film is optically very thick, i.e. very absorptive.  However, the reflectance curve of the capped reacted film  (Fig. \ref{fig:Fig_Optics} b) resembles the curve of the uncapped film (Fig. \ref{fig:Fig_Optics} c), that is, it exhibits interference fringes and a decrease of the reflection in the near infra-red. These results suggest that the as-deposited material is of metallic origin, which upon reaction with air turns into a wide-band gap semiconductor.  

The photochromic effect in an uncapped sample is illustrated in Fig. \ref{fig:Fig_Optics} d, where the optical transmittance before (clear state) and after 1 h illumination with a 3 mV, 405 nm violet laser (photodarkened state) are shown. The change in the transmittance is reversible, as shown in Fig. \ref{fig:Fig_Optics} e, where the average optical transmittance in the 500-800 nm range is depicted during 1h illumination/2 h darkness cycles.

\begin{figure}
\centering
\includegraphics[width=0.95\linewidth]{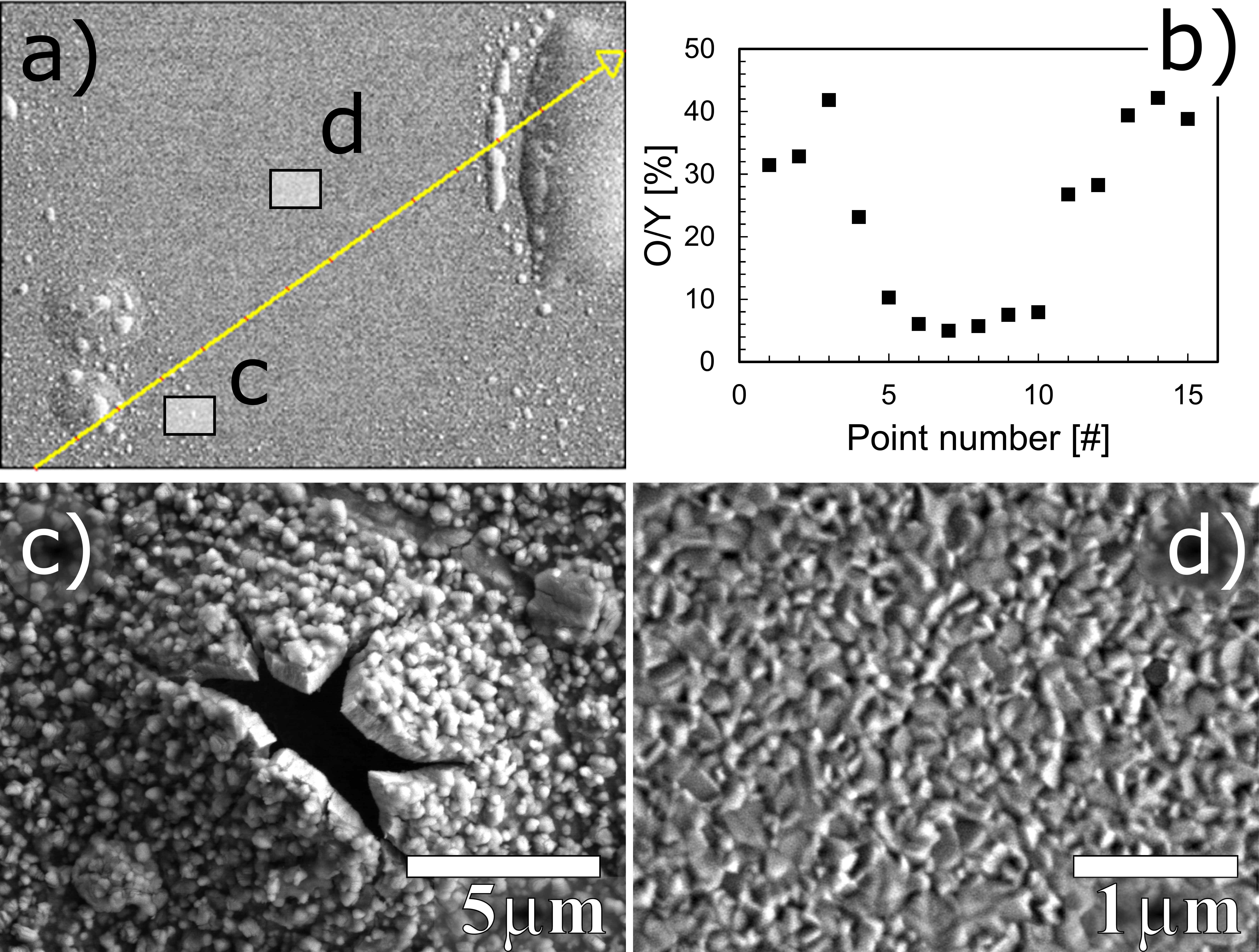}
\caption{Secondary electron micrographs and an EDS linescan of an aluminium capped YH$_{\mathrm{x}}$ film having partly reacted with air (a-d). The swollen left and right areas in micrograph (a) coincide with optically confirmed reacted areas, while the intact centre coincide with non-reacted areas. An EDS linescan (a-b) reveals that C$_{\mathrm{O}}$ $\approx$ 40 \% in the reacted parts of the YH$_{\mathrm{x}}$ (c) and  $\approx$ 5 \% in the non-reacted parts (d).}
\label{fig:SEM_Oxidation}
\end{figure}

Investigations with SEM and EDS were performed in order to gain insight into the reaction observed in the capped samples. In the secondary electron (SE) micrograph in figure \ref{fig:SEM_Oxidation}a, the combined film can be observed to have swollen and de-laminated in the areas to the left and right (Fig. \ref{fig:SEM_Oxidation}c), while being fully intact in the central areas (Fig. \ref{fig:SEM_Oxidation}d). By comparison with the visual appearance of the YH$_{\mathrm{x}}$ (similar to  Fig. \ref{fig:Oxidation2}b), the swollen, de-laminated areas were found to coincide with the areas of the YH$_{\mathrm{x}}$ visually confirmed to have undergone a reaction.  Upon investigation with EDS (a line scan from left to right - figure \ref{fig:SEM_Oxidation} a,b), the C$_{\mathrm{O}}$ in the reacted areas was measured as $\sim$40 \% at., approaching the $\sim$ 50 \% measured in uncapped samples. On the other hand, the non-reacted areas showed a C$_{\mathrm{O}}$ measuring $\sim$5 \%, a value that can be explained by for example a surface oxidation of the aluminium capping itself, a contribution from the glass substrate, and/or impurities in the microscope chamber. 

After argon gun sputtering through the capping layer, investigations with XPS showed a C$_{\mathrm{O}}$ in the non-reacted capped and reacted capped YH$_{\mathrm{x}}$ films measuring 18\% at. and 64\% at. respectively. The C$_{\mathrm{O}}$ in the pre-sputtered, non-capped, reacted YH$_{\mathrm{x}}$ was measured as 61\% at., being similar to that found for the reacted capped sample. The relatively high concentration of oxygen found in the non-reacted capped sample can be explained by post-sputter oxidation of the the topmost surface of the sample during the XPS measurement.  Contaminants in the XPS chamber can adsorb on the surface of the now unprotected YH$_{\mathrm{x}}$, yielding a significant signal in the surface sensitive XPS. This claim was proven by performing periodic measurements of the same sample which showed an increasing oxygen signal as a function of time. 

\begin{figure}
\centering
\includegraphics[width=1.0\linewidth]{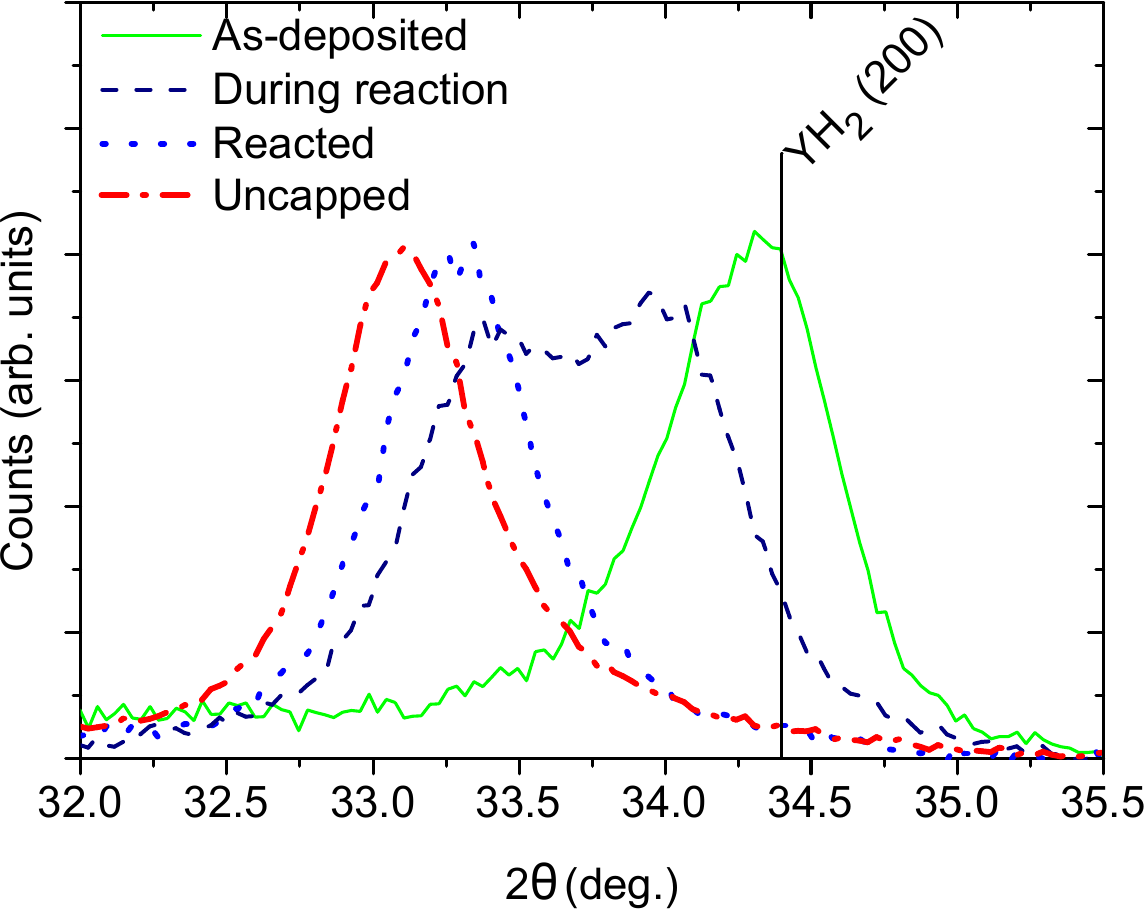}
\caption{The as-deposited YH$_{\mathrm{x}}$ forms in an fcc structure with a lattice parameter $a$ = 5.21 \AA \ (green). Upon reaction with air, the material retains the same symmetry but expands by approximately 3.5 \% to $a$ = 5.4 \AA \ (blue,red). During the reaction, the two phases coexist (purple), similar to what observed visually.}
\label{fig:XRD}
\end{figure}

Investigation of the non-reacted, capped material (Fig. \ref{fig:Oxidation2}a) using GIXRD reveals an fcc crystal structure and a lattice parameter $a$ = 5.2 \AA . The match between this and what has earlier been reported for yttrium di-hydride (YH$_{\mathrm{2}}$)\citep{Daou1} leads to the suggestion that this as-deposited material is in fact YH$_{\mathrm{2}}$, a claim is supported by the presented optical data. Upon exposure to air, the material was found to expand in the same fcc structure from an initial $a$ = 5.2 \AA \ to $a$ = 5.4 \AA \ . This transformation can be directly observed in the diffractograms in figure \ref{fig:XRD}, where the 200 reflection moves during the reaction, from the initial non-reacted state (green), via an intermediate state (purple), to the  fully reacted state (blue). In the intermediate state, the two extreme structures can be seen to co-exist, but with slightly distorted lattice parameters. It is noteworthy that the lattice parameter of the fully reacted state (blue) is similar to that previously reported for oxygenated yttrium hydride. \citep{Mongstad3,Maehlen1} The reacted, uncapped sample can in the same figure (red) be observed to have similar lattice parameter as the capped, reacted sample.

The lattice expansion caused by the introduction of oxygen might explain the observed swelling and de-lamination of the aluminium capped  YH$_{\mathrm{x}}$ (Fig. \ref{fig:SEM_Oxidation}c). A small oxygen permeability in the aluminium capping would be sufficient for the initiation of an oxidation reaction, and hence stress formation in the YH$_{\mathrm{x}}$. This stress could further cause an initial crack formation that would ensure even more incorporation of oxygen, eventually causing a cascade effect that would lead to the observed de-lamination.

Properties like band gap size and photodarkening potential are reported to be governed by the concentration of oxygen in YH$_{\mathrm{x}}$:O, \citep{You1} making control of the oxygen doping process highly important. As the oxygen in the reported fabrication method gets introduced after the formation of the metal hydride, it will largely be the properties of the metal hydride itself that determine the final degree of oxygen doping. Control of properties such as material porosity, grain structure and thickness are amongst others assumed to be of particular importance. 

In conclusion, we have found that photochromic oxygen-containing yttrium hydride can be prepared through an initial sputter deposition of oxygen free YH$_{\mathrm{x}}$ followed by a reaction with air. Optical inspections show that the as-deposited YH$_{\mathrm{x}}$ is dark and opaque, with metallic properties, while the oxygenated material is transparent and photochromic. Based on results obtained using XRD, the as-deposited YH$_{\mathrm{x}}$ is believed to be YH$_{\mathrm{2}}$ which upon reaction with oxygen expands and turns into the earlier reported photochromic oxygen containing yttrium hydride (YH$_{\mathrm{x}}$:O). \\\\

The authors would like to thank Dr. Chang Chuan You at the Institute for Energy Technology for fruitful discussions and support concerning sputter deposition. This work was financially supported by the Norwegian Research Council though the FRINATEK project 240477/F20 and the IDELAB/NANO2021 project 238848.


\end{document}